\documentclass[sigconf]{acmart}

\AtBeginDocument{%
  }

\copyrightyear{2026}
\acmYear{2026}
\setcopyright{cc}
\setcctype{by}
\acmConference[MSR '26]{23rd International Conference on Mining Software Repositories}{April 13--14, 2026}{Rio de Janeiro, Brazil}
\acmBooktitle{23rd International Conference on Mining Software Repositories (MSR '26), April 13--14, 2026, Rio de Janeiro, Brazil}
\acmPrice{}
\acmDOI{10.1145/3793302.3793377}
\acmISBN{979-8-4007-2474-9/2026/04}

\usepackage{subcaption}
\usepackage{tabularx}
\usepackage[dvipsnames]{xcolor}
\usepackage[normalem]{ulem}
\usepackage{enumitem}
\usepackage{pifont}
\usepackage{array}
\usepackage{lipsum}
\usepackage[utf8]{inputenc}
\usepackage{hyperref}
\usepackage{tabularray} 
\usepackage{graphicx}
\usepackage{caption}
\usepackage{tikz}
\usepackage{ragged2e}

\usepackage[most]{tcolorbox}
\usetikzlibrary{patterns}

\UseTblrLibrary{booktabs} 
\usetikzlibrary{calc}

\hypersetup{
    colorlinks=true,
    linkcolor=blue,
    filecolor=magenta,      
    urlcolor=cyan,
    pdftitle={Overleaf Example},
    pdfpagemode=FullScreen,
    }
\usepackage{amsmath}

\definecolor{mygrayzero}{gray}{0.9}
\definecolor{mygrayone}{gray}{0.8}
\definecolor{mygraytwo}{gray}{0.7}
\definecolor{mygraythree}{gray}{0.6}
\definecolor{Mycolor}{HTML}{166666}
\definecolor{revision}{HTML}{D10000}
\newboolean{showcomments}
\setboolean{showcomments}{true}
\ifthenelse{\boolean{showcomments}}
{
  \newcommand{\nbc}[3]{
    \colorbox{#3}{\bfseries\sffamily\scriptsize\textcolor{white}{#1}}
    {\textcolor{#3}{\sf\small$\blacktriangleright$\textit{#2}$\blacktriangleleft$}}
  }
}
{
  \newcommand{\nbc}[3]{}
}
\newcommand\functionalmark{\raisebox{-0.2em}{\includegraphics[width=1em]{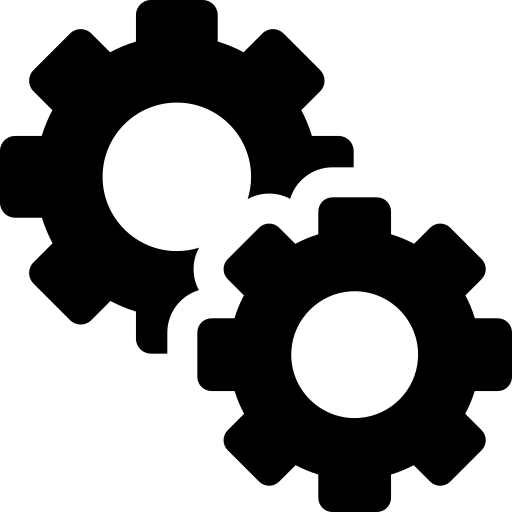}}} 
\newcommand\evolvemark{\raisebox{-0.2em}{\includegraphics[width=0.9em]{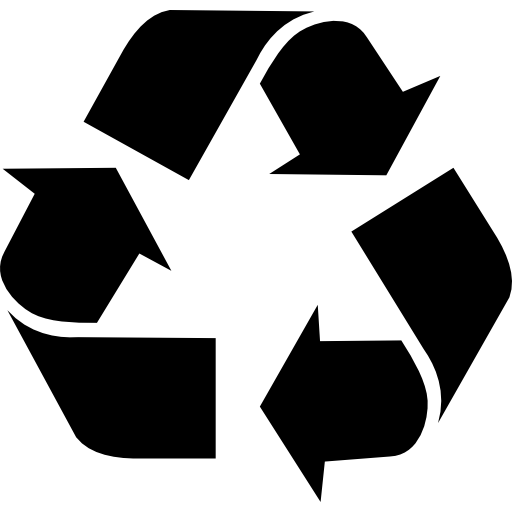}}} 
\newcommand\discussmark{\raisebox{-0.2em}{\includegraphics[width=1em]{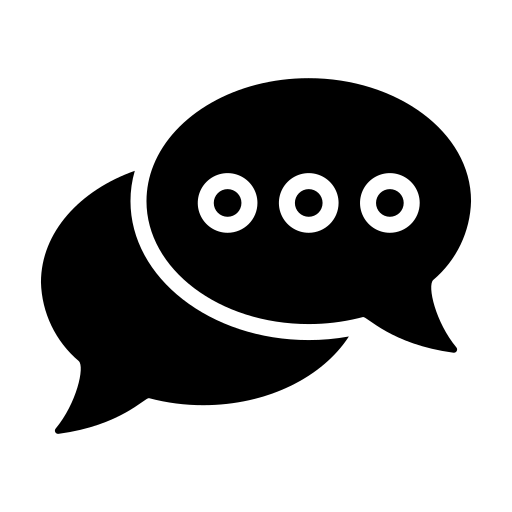}}}

\begin{document}

\title{Are We All Using Agents the Same Way? An Empirical Study of Core and Peripheral Developers’ Use of Coding Agents}
\author{Shamse Tasnim Cynthia} 
\affiliation{
    \institution{University of Saskatchewan} 
    \city{Saskatoon}
    \country{Canada}}
\email{shamse.cynthia@usask.ca}
\author{Joy Krishan Das}
\affiliation{
    \institution{University of Saskatchewan} 
    \city{Saskatoon}
    \country{Canada}}
\email{joy.das@usask.ca}
\author{Banani Roy}
\affiliation{
    \institution{University of Saskatchewan} 
    \city{Saskatoon}
    \country{Canada}}
\email{banani.roy@usask.ca}



\begin{abstract}

   Autonomous AI agents are transforming software development and redefining how developers collaborate with AI. Prior research shows that the adoption and use of AI-powered tools differ between core and peripheral developers. 
   However, it remains unclear how this dynamic unfolds in the emerging era of autonomous coding agents. In this paper, we present the first empirical study of 9,427 agentic PRs, examining how core and peripheral developers use, review, modify, and verify agent-generated contributions prior to acceptance. Through a mix of qualitative and quantitative analysis, we make four key contributions. First, a subset of peripheral developers use agents more often, delegating tasks evenly across bug fixing, feature addition, documentation, and testing. In contrast, core developers focus more on documentation and testing, yet their agentic PRs are frequently merged into the main/master branch. Second, core developers engage slightly more in review discussions than peripheral developers, and both groups focus on evolvability issues. 
   Third, agentic PRs are less likely to be modified, but when they are, both groups commonly perform refactoring. Finally,  peripheral developers are more likely to merge without running CI checks, whereas core developers more consistently require passing verification before acceptance.  Our analysis offers a comprehensive view of how developer experience shapes integration offer insights for both peripheral and core developers on how to effectively collaborate with coding agents.
    
\end{abstract}


\maketitle

\section{Introduction}\label{introduction}
AI-powered tools such as copilots and coding agents have significantly advanced AI-assisted software development, drawing significant attention from both industry~\cite{rondon2025evaluating} and academia~\cite{bouzenia2024repairagent, zhang2024autocoderover}. 
While copilots primarily provide chat-based assistance that guides developers as they make changes within their own development environment,  coding agents put forward greater autonomy.  
These agents operate within their ephemeral development environment, execute and adjust their actions based on feedback from compilers or linters to autonomously resolve issues and submit pull requests (PRs). This promises faster delivery cycles \cite{bird2022taking} and reduced cognitive load~\cite{stray2025generative}, leading to their growing adoption by developers across all experience levels ~\cite{watanabe2025use, thomas}. 

Despite increasing agent adoption, prior research has primarily examined chat-based assistance such as Copilot and ChatGPT. For instance, Barke et al. \cite{barke2023grounded} showed that developers use Copilot either to speed up or to explore options when unsure. Similarly, a study with student programmers~\cite{prather2023s} found that despite challenges interpreting Copilot’s suggestions, participants remained optimistic about its future use. These studies, in their early forms, highlight how developers feel about AI-assistance. While these studies offer useful insights, we still lack empirical evidence on how different types of developers interact with autonomous coding agents that behave more like fellow engineers~\cite{chen2025code}.

This understanding is particularly important as most assessments on coding agents stem from benchmarks like SWE-Bench~\cite{jimenez2024swebench, zhang2024autocoderover, xia2024agentless, yang2024swe}.  
Furthermore, some prior studies have shown that developers’ perceptions and adoption patterns of AI-powered tools vary significantly with experience.
For instance, Peng et al.~\cite{peng2023impact} showed that peripheral developers completed tasks 55.8\% faster with Copilot in a controlled study. Conversely, McKinsey~\cite{mckinsey} and Jellyfish~\cite{jellyfishAnalyzed146000} found that core developers benefited more substantially compared to junior developers (e.g., 7-10\% more time to complete tasks), and Vaithilingam et al.~\cite{vaithilingam2022expectation} observed no significant experience-based differences. Collectively, these contrasting findings reveal that developer experience plays
a critical role in shaping how AI-powered tools are adopted and utilized. However, we still lack empirical evidence on how this dynamic unfolds between core and peripheral developers when working with autonomous coding agents in real-world workflows.

To address this gap, we conduct an empirical study of $9,427$ resolved agentic PRs on GitHub attributed to four widely used agents (Claude, Copilot, OpenAI Codex, and Cursor). Our aim is to understand how collaboration with agents differs between core and peripheral developers across the pull request (PR) lifecycle (i.e., from submission to acceptance). We proceed in four steps. First, we quantify how often developers delegate tasks to agents and for what purposes. Second, we assess how intensely agent-generated code is reviewed and what issues are raised. Third, we measure how much and what types of developer modifications are applied to agent code. Last, we evaluate to what extent the merge verification check passes at acceptance. Our analysis offers a comprehensive view of how developer experience shapes delegation, review, modification, and verification, providing practical guidance for collaborating with agents to achieve reliable contributions across experience levels.

This paper makes four major contributions by answering four research questions as follows. 

\begin{itemize}[leftmargin=*] 
 \renewcommand\labelitemi{\ding{43}}
    \item \textbf{RQ\textsubscript{1}.}~\textbf{To what extent do core and peripheral developers differ in the frequency and purpose of agent use on GitHub?} Although trust is inherently subjective, prior studies suggest that it can be inferred from an individual’s usage of a system~\cite{baek2023chatgpt, kim2019study} and the nature of tasks they are willing to delegate~\cite{sheth1991we}. To investigate how these behaviors vary across core and peripheral developers, we analyze $9,427$ closed agentic PRs. We find that, although usage rates are similar, a subset of peripheral developers uses agents disproportionately more often. Finally,  peripheral developers delegate tasks evenly across \textit{bug fixing}, \textit{feature addition}, \textit{documentation}, and \textit{testing}),  whereas core developers focus more on \textit{documentation} and \textit{testing}.

    \item \textbf{RQ\textsubscript{2}.}~\textbf{How do core and peripheral developers differ in the intensity of code review and the types of issues they raise on agentic PRs?} Rigorous code review is known to improve software quality~\cite{mcintosh2014impact}. We therefore examine how the intensity of review comments and the types of issues raised on agentic PRs differ between core and peripheral developers. We find that core developers engage slightly more in review discussions than peripheral contributors. Both groups primarily address evolvability-related issues in agent-generated code. Within this category, peripheral developers tend to focus more on \textit{code organization}, while core developers more frequently prioritize \textit{alternative solution} approaches.

    \item \textbf{RQ\textsubscript{3}.}~\textbf{How do core and peripheral developers differ in the intensity and types of modifications applied to agent-gene-rated code?}
    After developer's review (RQ\textsubscript{2}), agents may apply changes. However, it remains unclear whether developers still intervene to ensure that the agent's code adheres to project-specific standards. Therefore, we analyze how core and peripheral developers modify agentic PRs, measuring both the intensity and types of modifications made to agent's code. We find that, on average, $74.1\%$ of agentic PRs are accepted without modification across both groups. When modifications do occur, both core and peripheral developers commonly \textit{refactor agent-generated code}. 

    \item \textbf{RQ\textsubscript{4}.} \textbf{What differences exist in verification outcomes for agentic PRs accepted by core versus peripheral developers?}
    Modern Code Review (MCR)~\cite{mcintosh2014impact, tsay2014let} is increasingly paired with automated verification checks through Continuous Integration (CI) pipelines. These checks run automatically on each change, acting as a gate during acceptance or merge. To examine whether experience shapes integration outcomes, we analyze how often agentic PRs from core and peripheral developers successfully pass all verification checks prior to acceptance. We find that peripheral developers are more likely to merge agentic PRs without running any checks, whereas core developers are more likely to ensure that accepted agentic PRs have passed the CI pipeline.

\end{itemize}

\noindent \textbf{Replication package} including the scripts and data to answer our RQs can be found in our online appendix \cite{replication_package}.

\begin{figure*}
    \centering
    \resizebox{0.8\textwidth}{!}{
    \includegraphics[width=0.8\textwidth]{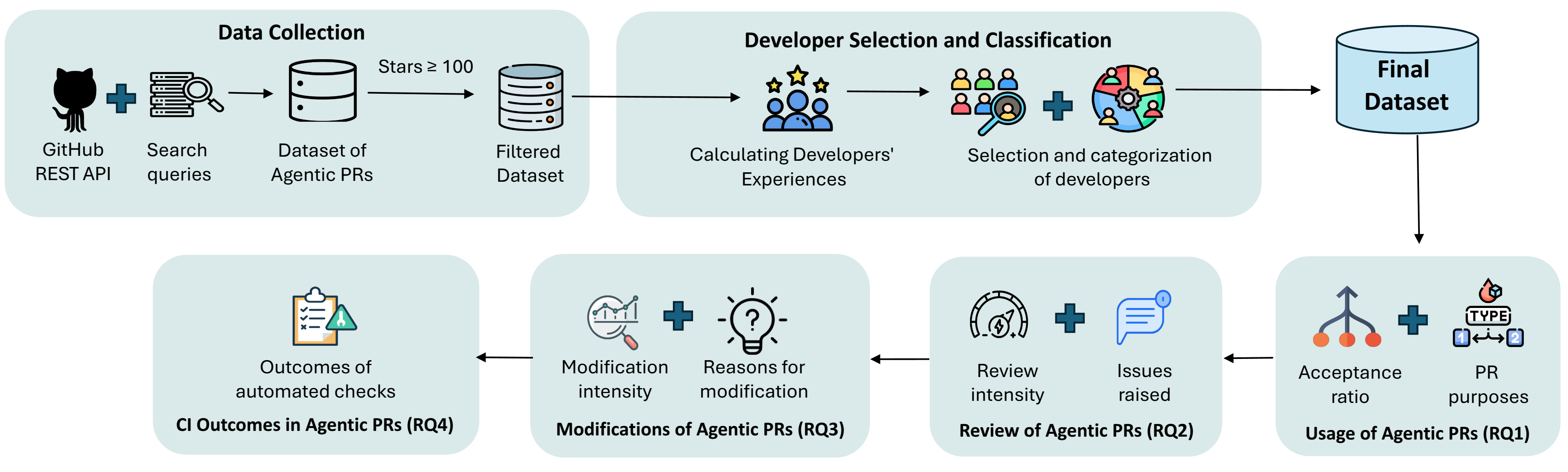}
    }
    \caption{Overview of the methodology to investigate core and peripheral developers' use of coding agents}
    \label{fig:methodology}
\end{figure*}

\section{Methodology}
Figure~\ref{fig:methodology} demonstrates the methodology of our study. The following sections provide a detailed description of each method.

\subsection{Data Preparation} \label{subsec:Data-collection}
In this section, we describe the data preparation procedure.

\begin{table}[t]
    \centering
    \caption{GitHub search queries to extract PRs following \cite{li2025rise}}
    \resizebox{0.9\linewidth}{!}{
    \begin{tabular}{
      >{\raggedright\arraybackslash}p{2cm} 
      p{4cm}       
      r                              
    }
    \toprule
        Agent       & GitHub Search Query & Release Date \\
    \midrule
    Claude Code     &   is:pr ``Co-Authored-By: Claude" &   2025-02-24 \\
    Cursor          &   is:pr head:cursor/              &   2025-01-01 \\
    GitHub Copilot  &   is:pr head:copilot/             &   2025-01-01  \\
    OpenAI Codex    &   is:pr head:codex/               &   2025-05-16  \\
    \bottomrule
    \end{tabular}
    }
    \label{tab:pr-queries}
\end{table}

\textit{Data Collection.} We extracted PRs from GitHub associated with four widely adopted coding agents: Claude Code, Cursor, GitHub Copilot and OpenAI Codex. These agents were selected due to their increasing adoption in practice and their superior performance compared to other coding agents on the SWE-Bench\footnote{\url{https://www.swebench.com/}}. Data collection was conducted using the GitHub REST API\footnote{\url{https://docs.github.com/en/rest?apiVersion=2022-11-28}} in combination with targeted search queries, following a methodology similar to Li et al. \cite{li2025rise}. Table~\ref{tab:pr-queries} presents the specific search queries employed to identify PRs linked to each agent. These queries were adapted from PRarena\footnote{\url{https://github.com/aavetis/PRarena}} and leverage distinctive indicators such as branch name prefixes (e.g., ``head:copilot/") or attribution statements in the PR body (e.g., ``Co-Authored-By: Claude"). We further customized the search to include a specialized query for Claude Code and applied a start-date filter to remove unrelated PRs. Our dataset includes data up to August 1, 2025. Moreover, to ensure project relevance and data quality, we applied a filter to include only repositories with $\geq$100 GitHub stars, following the best practices for GitHub-based empirical analyses~\cite{ kalliamvakou2014promises}. Finally, our dataset included $28,769$ PRs, with $2,856$ developers from $2,595$ repositories. In the subsequent step, we describe how developer experience was measured.

\textit{Developer Experience.} 
Prior literature establishes that a developer's project-specific experience is strongly correlated with their contribution volume~\cite{posnett2013dual, mockus2010organizational}. As developers contribute to a project's codebase through various activities, they transition from peripheral contributors to core maintainers ~\cite{crowston2006core}.
Researchers have measured developers' experience to categorize them into core or peripheral using various contribution metrics. This includes the number of commits \cite{eyolfson2011time, posnett2013dual}, lines of code contributed \cite{mockus2002two}, and developer seniority (i.e., years since first code contribution) \cite{mockus2010organizational}. Among these, commit count has been the most common proxy for developer experience~\cite{eyolfson2011time}. 

However, commit-based measures have known limitations. Commits capture a single logical unit of work and are influenced by individual development styles~\cite{robbes2013using}. For instance, one developer may commit frequently in small increments, whereas another may batch equivalent changes into fewer commits, leading to skewed experience estimates. Moreover, when developers transition to core roles, their commit activity often decreases as they shift toward coordination and review responsibilities~\cite{jergensen2011onion}.

Thus, prior work recommends estimating developer experience using a broad set of contribution activities~\cite{argote2011organizational, robbes2013using}. Following this guidance and recognizing pull-based development has become the dominant paradigm in open source software (OSS) projects~\cite{gousios2014exploratory}, we adopt a PR-based metric to calculate developers' experience, as each PR encompasses both code commits and review~\cite{dey2020effect}. We measured a developer’s experience within a repository as the normalized sum of authored and reviewed closed PRs relative to the total number of closed PRs, following Kononenko et al.~\cite{kononenko2016code}.

\textit{Developer selection.} To retrieve a developer's activity, we used GitHub’s REST API and their unique \texttt{login} identifiers. 
Next, we ensured that the experience metric only reflects preexisting expertise rather than PR activity induced by coding agents. Including developers whose contribution began after the introduction of agents would confound whether experience influenced agent usage or vice versa. Thus, we filtered out developers who had not authored any PRs before the introduction of agents (i.e., before January 2025 \cite{li2025rise}). After this step, we identified $2,057$ unique developers associated with $12,925$ PRs and $1,899$ repositories.  

Since a developer can contribute to multiple repositories, experience scores may differ across projects. 
In prior work, developer roles are typically defined at the project level \cite{campos2022empirical}. 
Thus, to avoid role ambiguity, where a developer could be core in one repository but peripheral in another, we further removed developers who collaborated on agentic PRs in more than one repository.
Moreover, as our study targets completed development activities, we excluded all open PRs. 
Therefore, after applying these filtering steps, our final dataset consisted of $9,427$ PRs from $1,391$ repositories and $1,701$ developers, whose activity was retrieved, and for whom we calculated their corresponding experience scores.

\textit{Developer Classification.} 
We classify developers into two groups (i.e., core and peripheral) based on their computed experience scores. Prior research ~\cite{crowston2006core, mockus2002two} consistently shows that software projects exhibit skewed contribution patterns, where few developers account for the majority of contributions (i.e., approximately 80\% ~\cite{mockus2002two}), while most contribute relatively little. In accordance, \textit{core} developers are defined as those who contribute the most to a project, while the others are \textit{periphery}. Following this established convention, we adopt the 80\textsuperscript{th} percentile of the experience score distribution as the threshold for distinguishing core contributors from peripheral ones. With this criterion, we identify 390 developers as core contributors, while the remaining 1,311 developers are classified as peripheral contributors. 
Table~\ref{tab:dataset-summary} shows the descriptive statistics about our corpus after the data preparation steps have been applied. 
\begin{table}[t]
  \centering
  \caption{An overview of the subject dataset}
  \label{tab:dataset-summary}
  \begin{tabular}{lr}
    \toprule
    \textbf{Item} & \textbf{Value}\\
    \midrule
    \# of total closed/resolved agentic PRs            & 9,427 \\
    \# of repositories                        & 1,391 \\
    \# of merged PRs (\textit{accepted})      & 6,819 \\
    \# of developers & 1,701 \\
    \# of core developers       & 390  \\
    \# of peripheral developers  & 1,311  \\
    \bottomrule
  \end{tabular}
\end{table}

\subsection{Characterizing Agentic PR Usage} \label{subsec:RQ1-usage}
To answer RQ\textsubscript{1}, we first measure and compute the frequency of accepted PRs among the $9,427$ PRs across the core and peripheral groups of developers. We employ metrics and perform statistical analysis that have been previously applied in the literature \cite{thongtanunam2016revisiting}. 

Second, we aim to characterize the primary purpose of agentic PRs across different developer groups. Here, we limit our dataset to 6,819 accepted PRs to focus exclusively on contributions with the developer's intent to merge. Given the emerging stage of agent adoption and the low barrier to entry, developers explore agents' capabilities by prompting them to resolve hypothetical issues~\cite{subramonyam2025prototyping} rather than assigning actual issues. These exploratory PRs often lack substantive developer engagement, such as code reviews or follow-up modifications. For instance, in \textit{PR\#699}\footnote{\url{https://github.com/Aiko-IT-Systems/DisCatSharp/pull/699}}, a developer prompted the Copilot agent to draft an experimental abstraction hierarchy but subsequently cancelled the operation after observing the initial output, and then closed the PR. Therefore, by focusing on accepted PR, we isolate our study to PRs that resolve real-world issues within the project codebase.

Since a PR consists of one or more interrelated commits~\cite{liu2019automatic}, we adopted the ten-category taxonomy proposed by Zeng et al.~\cite{zeng2024first}, which was originally developed for commit-level classification. However, we applied this taxonomy to characterize the purpose of agentic PRs, following a similar strategy to prior work~\cite{watanabe2025use}. The taxonomy categories are outlined in Table~\ref{tab:pr-purposes}.

\begin{table}[th]
    \centering
    \vspace{-0.2cm}
    \caption{Taxonomy used for categorizing PR purposes}
    \resizebox{1\columnwidth}{!}{
        \begin{tabular}{p{1.3cm}p{10cm}}
            \toprule[1pt]
            \textbf{Category} & \textbf{Description} \\ 
            \midrule
            fix & Bug and fault correction in the codebase. \\ \midrule
            feat & Code changes that introduce new functionalities, including internal or user-facing features. \\ \midrule
            refactor & code restructuring to improve maintainability without altering behavior. \\ \midrule
            docs & Updates documentation, or comments (e.g., README, API docs, typo fixes). \\ \midrule
            test & Adds or modifies test cases or test files. \\ \midrule
            build & Code changes that modify build systems or dependencies (e.g., Maven, Gradle, Cargo). \\ \midrule
            style & Enhances code readability or formatting without changing behavior. \\ \midrule
            ci & Updates CI/CD workflows and configuration files. \\ \midrule
            perf & Improves execution speed, memory usage, or efficiency. \\ \midrule
            chore & General maintenance tasks (e.g., dependency bumps, cleanup). \\ 
            \bottomrule
            \multicolumn{2}{l}{\small Note: We adopt the taxonomy from established prior work~\cite{zeng2024first}.}
        \end{tabular}
    }
    \label{tab:pr-purposes}
\end{table}
\vspace{-0.3cm}
To characterize the purpose of the $6,819$ accepted PRs, we employed the GPT-4 model to identify the PR categories from the taxonomy, following a recent SE study ~\cite{kumar2025time}. We selected GPT-4 due to its demonstrated capability in accurately annotating and categorizing SE artifacts~\cite{ahmed2025can}. To ensure reliability, we adopted a human validation approach consistent with prior work~\cite{kumar2025time}. 
Specifically, we selected a random stratified sample of 364 PRs with proportionate strata (satisfying 95\% confidence level with a 5\% error rate). Here, proportionate stratification is appropriate because the goal is to validate the labeling accuracy of GPT-4 over the entire dataset, not to compare performance across strata. 
Then, we used GPT-4 to assign category labels from the selected taxonomy based on the PR title and body, allowing multiple labels per PR when applicable~\cite{watanabe2025use}. Next, the first two authors independently reviewed GPT-4’s annotations for this subset, adding alternative labels where appropriate and documenting disagreements. Cases with inconsistent labeling were flagged for further review and underwent secondary evaluation with refinement in the prompting strategies \cite{kahol2025oss}. The resulting inter-rater reliability (IRR), measured using Cohen’s Kappa, was 80.3\%, indicating strong agreement among annotators~\cite{mchugh2012interrater}. Given this validation and the higher observed alignment with GPT-generated labels, the remaining dataset was classified using GPT-4.


\subsection{Analyzing Review Dynamics in Agentic PR} \label{subsec:RQ2-reviews}

We aim to understand how rigorously agentic PRs are reviewed, as well as the patterns of issues raised during the review process by both core and peripheral developers, in order to address RQ\textsubscript{2}. 
Therefore, we analyze the $6,819$ accepted PRs to identify those with review comments.  We found $1,501$ PRs containing review comments, resulting in $9,403$ comments collected via GitHub's REST API endpoint\footnote{https://api.github.com/repos/\{owner\}/\{repo\}/pulls\{pr-number\}/comments}. We perform quantitative analysis on these comments to investigate the review intensity.

\begin{table}[t]
    \centering
    \caption{Review comments taxonomy and description}
    \resizebox{1\columnwidth}{!}{
        \begin{tabular}{p{3.5cm}p{10cm}}
            \toprule[1pt]
            \textbf{Category} & \textbf{Description} \\ 
            \midrule
            \functionalmark~Functional Defect & A functionality is missing or implemented incorrectly, which often requires additional code or major fixes. \\ \midrule
            \functionalmark~Validation & Issues with invalid value detection and data sanitization. \\ \midrule
            \functionalmark~Logical & Issues involving comparison, control flow, computation, or other logical errors. \\ \midrule
            \functionalmark~Interface & Issues arising from external components such as libraries, hardware devices, database, or operating system. \\ \midrule
            \functionalmark~Resource & Issues related to the initialization, handling, or release of variables, memory, files, and databases. \\ \midrule
            \functionalmark~Support & Issues related to support systems, libraries, or their configurations. \\ \midrule
            \functionalmark~Timing & Issues caused by improper thread synchronization when accessing shared resources. \\ \midrule
            \evolvemark~Solution Approach & Suggestions for alternate implementations (e.g., algorithms, data structures). \\ \midrule
            \evolvemark~Documentation & Suggestions to improve code comments or documentation. \\ \midrule
            \evolvemark~Organization of Code & Suggestions for structural refactoring, such as collapse hierarchy, extract superclass, and inline function. \\ \midrule
            \evolvemark~Alternate Output & Suggestions for enhancing error messages, alerts, toast notifications, or function return values. \\ \midrule
            \evolvemark~Naming Convention & Suggestions for renaming software elements to comply with conventions. \\ \midrule
            \evolvemark~Visual Representation & Suggestions for enhancing code readability, such as adjusting indentation, removing unnecessary whitespace, or reorganizing code. \\ \midrule
            \discussmark~Question & Questions to understand design and implementation choices. \\ \midrule 
            \discussmark~Design Discussion & High-level discussion about design choices, design patterns, and software architecture. \\ 
            \bottomrule[1pt]
            \multicolumn{2}{l}{\small Note: We adopt the code review comment categories and their descriptions from established prior work.~\cite{beller2014modern, mantyla2008types, turzo2024makes}}\\
            \multicolumn{2}{l}{\small \functionalmark~Functional, \evolvemark~Evolvability, \discussmark~Discussion}\\
        \end{tabular}
    }
    \label{tab:review-taxonomy}
\end{table}

Next, to identify the patterns of issues raised by developers during these reviews, we apply a closed coding analysis~\cite{krippendorff2018content}. We take the set of labels proposed by prior studies~\cite{beller2014modern, mantyla2008types} as our code booklet (Table~\ref{tab:review-taxonomy}). 
After this, we followed Tufano et al. \cite{tufano2021towards} to preprocess the review comments. Specifically, we use the \texttt{code/Cleaner.py} script from their provided replication package. The script identifies non-English comments using multiple detectors and only retains comments classified as English. Furthermore, comments consisting solely of emojis, hyperlinks, acknowledgements (e.g., \textit{``LGTM"}, \textit{+1}), or conversational replies (e.g., \textit{``see above"}, \textit{``same here"}) are removed, as they do not convey actionable review intent. On top of that, we excluded comments authored by bots or agents, whose username contained the suffix \textit{``bot''} \cite{golzadeh2022accuracy}, and accounts matching an established list of known bots or coding agents \cite{golzadeh2021ground}.
Subsequently, we also removed the comments made by other developers who were not present in our studied $1,701$ developers. 
This resulted in a total of $2,735$ comments from both developer groups.

Then, we employed stratified random sampling with equal strata, resulting in a total sample of $340$ comments (satisfying a $95\%$ confidence level and a $\pm$5\% margin of error). 
As our objective is to characterize patterns of each developer group independently rather than to analyze for the aggregate population, we applied group-specific sampling and analysis strategies following prior work ~\cite{guo2024exploring}.
The coding task was performed by the first two authors as coders. Initially, the coders independently labeled a separate set of 40 comments, achieving a Cohen's kappa coefficient of 0.71 (moderate agreement). Since this score was not sufficient to proceed, the authors convened to discuss the labeling. They addressed disputes and conflicts until reaching a consensus. After refining the labels, they re-labeled the 40 samples and confirmed their agreement. The IRR, measured by Cohen’s Kappa, was found to be 84.5\%, indicating near-perfect agreement~\cite{mchugh2012interrater}. 
Finally, the first two authors labeled the same set of comments individually and achieved a Cohen's Kappa score of 82.5\%, signifying strong agreement~\cite{mchugh2012interrater}.

\subsection{Characterizing Agentic PR Modifications}\label{subsec:RQ3-modifications}
To answer RQ\textsubscript{3}, we filter the agentic PRs that were subsequently modified by developers from the total of $6,819$ accepted PRs. We define an agentic PR as modified if a developer added one or more commits after the agent's initial submission. Although agents may also modify a PR in response to feedback, our study's focus is on understanding whether the degree and purpose of developer-driven modifications differ between core and peripheral contributors.  Therefore, we included only those PRs in which the developer made subsequent commits following the agent’s contribution. 
To do so, we first collect commit-level data using the GitHub REST API endpoint\footnote{https://api.github.com/repos/\{owner\}/\{repo\}/commits/\{commit-sha\}} for all $6,819$ accepted PRs. 
%
Next, we identify and distinguish developers' commits from the agent's commits within these PRs. When agents commit, they compose the commit message themselves and often advertise their presence in several ways: (1) By being the \textit{author} of the commit, analogous to other bots (e.g., Dependabot). For example, commits by Copilot list the author as ``copilot-swe-agent[bot]'' with the login ``Copilot''. Similarly, commits by Cursor use the author name ``Cursor Agent'', the email ``cursoragent@cursor.com'' and the login ``cursoragent''. 
(2) By adding themselves as a \textit{co-author} of the commit, with the agent's user recorded as the main author. For instance, commits by Claude used the ``Co-authored-by:'' trailer added. (3) By adding an additional trailer (such as \textit{label}) when authoring a commit event. For example, a commit event by Codex had a ``Codex'' label added prior to it.
Any commit that did not match the above patterns was classified as \textit{developer-authored} commit. This filtering process enabled us to isolate $4,339$ agent-assisted PRs where developer modifications were present. On these PRs, we measure and compare the degree of modifications between the developer groups based on lines of code added and deleted.

Then, to further understand the reasons behind these modifications, we conducted a closed coding analysis. As done in Section~\ref{subsec:RQ2-reviews}, we randomly sample $360$ commit messages (satisfying a 95\% confidence level with a ±5\% margin of error).
We used the commit-level taxonomy proposed by Zeng et al.~\cite{zeng2024first} as our booklet. Following prior work~\cite{jiang2021developers, nejati2023code}, commits with ambiguous descriptions were excluded and replaced with new samples to maintain the target sample size.
The coding was performed by the first two authors, who had already established a shared understanding of labeling criteria through the earlier coding process (Section~\ref{subsec:RQ1-usage}). Thus, both coders independently annotated the commits, achieving an IRR (Cohen’s Kappa) of 81.9\%, indicating strong agreement between annotators~\cite{mchugh2012interrater}. Any disagreements that occurred were resolved through discussions.

\subsection{Analyzing CI Outcomes in Agentic PR}
Modern Code Review (MCR)~\cite{mcintosh2016empirical} is increasingly coupled with automated quality checks (i.e., static analysis, test execution, and build validation)~\cite{hilton2017trade}. Continuous integration (CI) features the automated execution of these checks every time a change is pushed. These self-testing safeguards aim to detect defects early and ensure that only secure and reliable changes enter a project's branch~\cite{fowler2006continuous}. Importantly, all CI jobs should pass during the merge stage~\cite{soares2022effects} and a previous study highlights that PRs with more passed checks have a higher chance of being merged than broken CI~\cite{zampetti2019study}.
Since agentic PRs are increasingly merged into the project codebase~\cite{li2025rise}, we aim to examine how often these automated checks are bypassed by core and peripheral developers during merge commit, answering RQ\textsubscript{4}.

In GitHub, whenever a new commit is added to a PR (by a developer or an agent), the platform automatically triggers \texttt{check\_suite} event with an action of \texttt{requested} to all GitHub Apps installed by the project that have write permission. Note that a \texttt{check\_suite} is a collection of the check runs created by a single CI App in GitHub, which summarizes the status and conclusion of the check runs. We use the GitHub REST API endpoint\footnote{https://api.github.com/repos/\{owner\}/\{repo\}/commits/\{sha\}/check-runs} to collect the \texttt{completed} status of $6,819$ accepted agentic PRs and compare the conclusion across the developer groups. When a check run completes, its conclusion can be \textit{success}, \textit{failure}, \textit{neutral}, \textit{cancelled},\textit{ timed\_out}, \textit{skipped}, or \textit{action\_required}. 

Based on the conclusion, we determine the rollup results of the CI checks, following a prior study~\cite{khatami2024state}. That is, if all checks' conclusion is ``success'', then the rollup status of the CI is success, otherwise it is failure.
Moreover, to ensure that the results accurately reflected the actions of our studied developers, we included only those PRs where the same developer both assigned the task to the agent and merged it. This filtering resulted in a final analysis set of 5,319 PRs from peripheral developers and 1,018 PRs from core developers to investigate the rollup results of CI checks.


\section{Study Findings}
We ask four research questions in this study. In this section, we answer them carefully with the help of our quantitative and qualitative findings as follows:

\subsection{Usage Characteristics of Agentic PR (RQ\textsubscript{1})} \label{result-RQ1}

\begin{figure}[t]
    \centering
    \includegraphics[width=\linewidth]{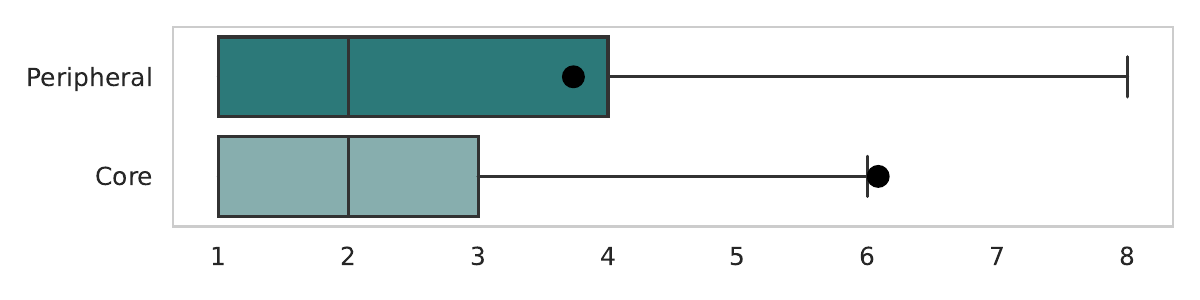}
    \caption{Distribution of resolved PRs across core and peripheral developers ($\bullet$ represents mean here)}
    \label{fig:closed-PRs}
\end{figure}

Figure~\ref{fig:closed-PRs} provides a distribution of resolved agentic PRs across core and peripheral developers.  Although the median number of agent usage per developer is identical across both groups ($2.0$), the mean is substantially higher for peripheral developers ($6.08$) compared to core developers ($3.73$). This suggests that while the \textit{regular} level of AI delegation is similar, a subset of peripheral contributors generates disproportionately more agentic PRs. Our \textit{Mann-Whitney Wilcoxon} \cite{fay2010wilcoxon} test also revealed a statistically significant difference ($p-value<0.05$) between these two groups. However, the Cliff's Delta size is negligible ($|d|=0.06$). These results suggest that, though statistically significant, the overall resolution patterns of agentic PRs between the two groups have minimal practical impact.

Moreover, as shown in Table~\ref{tab:created-merged-PRs}, the overall acceptance rates of agentic PRs appear similar for core and peripheral developers. However, for the accepted PRs that were merged into the \texttt{main}/\texttt{master} branch, core developers exhibit a substantially higher merge rate than peripheral contributors. Prior research has shown that developer reputation within a repository can strongly influence merge outcomes ~\cite{wang2023fork}, with PRs from core contributors being more likely to reach the main branch compared to those from peripheral developers~\cite{ortu2020you}.

\begin{table}[t]
    \centering
    \caption{PR acceptance ratio across core and peripheral developers}
    \resizebox{1\linewidth}{!}{
    \begin{tabular}{p{1.5cm}rrrrr}
    \toprule
        Developer Category & \#Devs & \#Resolved  & \#Accepted &  Accepted(\%)  & \begin{tabular}{@{}c@{}}~$\mapsto$m(\%)\end{tabular}\\
    \midrule
        Peripheral   & 1,311 & 7,971 & 5,759 & 72.3  & 77.8  \\
        Core         & 390 & 1,456 & 1,060 & 72.8  & 85.8  \\
        
    \bottomrule
    \multicolumn{6}{l}{\small  ``$\mathbf{\mapsto}$~m'' signifies PR been accepted or merged into the main/master branch.}\\
    \end{tabular}}
    \label{tab:created-merged-PRs}
    \vspace{-0.5cm}
\end{table}

\begin{figure}[!t]
    \begin{subfigure}{0.9\linewidth}
        \centering
        \includegraphics[width=\linewidth]{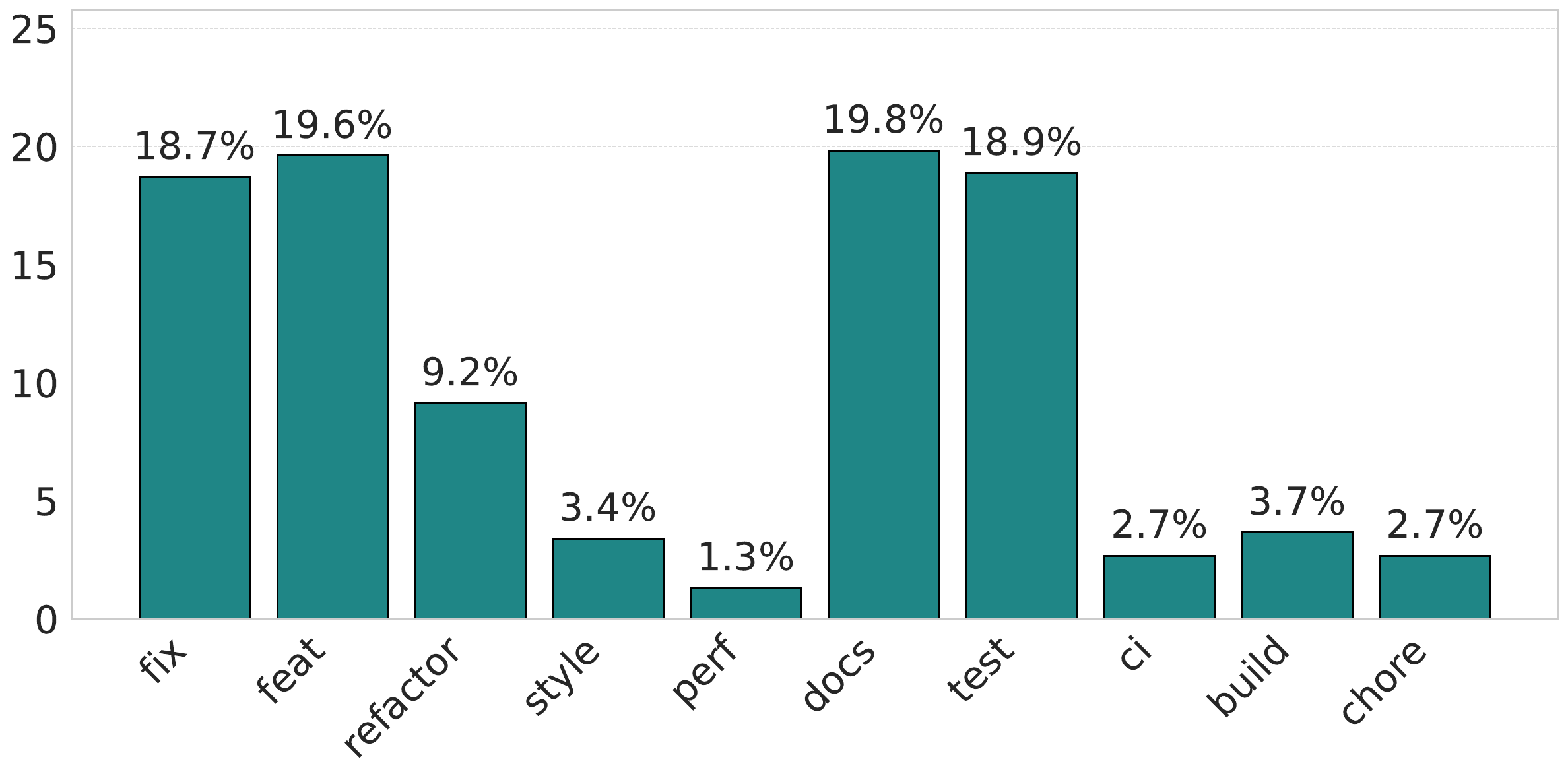}
        \caption{Peripheral developers}
        \label{fig:peri_pr}
    \end{subfigure}
    \begin{subfigure}{1\linewidth}
        \centering
        \includegraphics[width=0.9\linewidth]{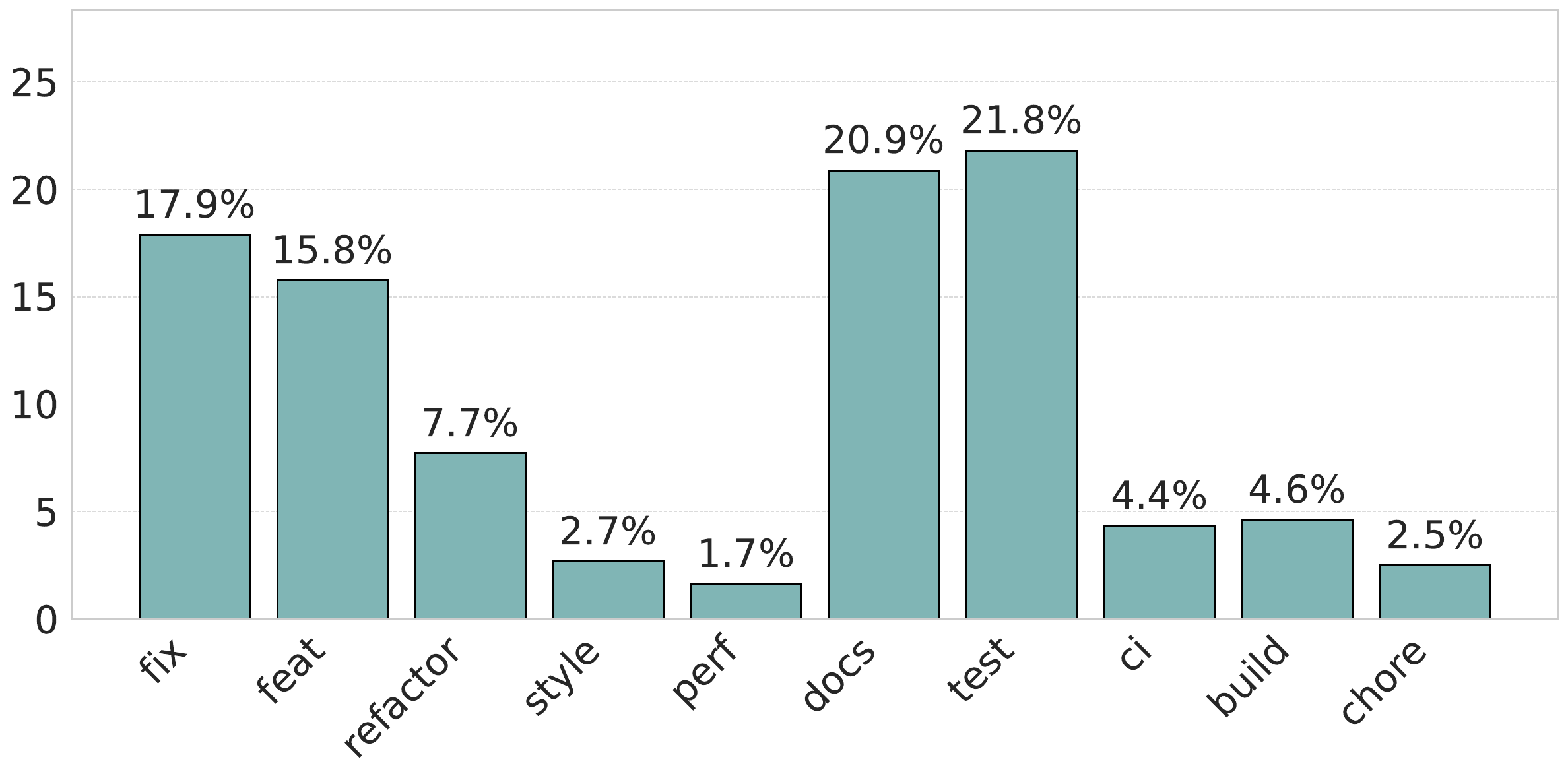}
        \caption{Core developers}
        \label{fig:core_pr}
    \end{subfigure}
    \caption{Purposes of agentic PRs across core and peripheral developers}
    \label{fig:pr_labels}
    \vspace{-0.5cm}
\end{figure}

Next, Figure~\ref{fig:pr_labels} represents the underlying purposes of agentic PRs created by each developer group. Both core and peripheral developers use agents across all \textit{ten} purpose types. However, agentic PRs from both groups were mainly concentrated in \textit{four} primary categories: \textit{documentation (docs)}, \textit{testing (test)}, \textit{feature implementation (feat)}, and \textit{bug fixing (fix)}. For core developers, these four categories accounted for $76.4\%$ of their accepted agentic PRs, with the remaining PRs spread across \textit{refactoring (refactor)}, \textit{non-functional adjustments (style)}, \textit{performance improvements (perf)}, \textit{CI/CD updates (ci)}, \textit{build configuration changes (build)}, and \textit{other cleanup activities (chore)}. Peripheral developers show a similar pattern,  where $77\%$ of their agentic PRs fall within the four primary categories.

Moreover, \textit{\textbf{peripheral developers showed an even pattern of agent use across the four major PR types}}\textit{(\textasciitilde$18.7$–$19.8\%$)}. Our thematic analysis further reveals that peripheral developers' activities showed a uniform AI delegation strategy rather than category-specific reliance. 
For example, in \textit{PR\#86}\footnote{\url{https://github.com/Deep-Learning-Profiling-Tools/triton-viz/pull/86}}, the developer delegated several tasks to the agent. The agent fixed multiple bugs in both the frontend and backend that were preventing the visualization system from loading and displaying data correctly. It also added new features such as clearer server messages. In addition, the agent updated documentation by adding new example scripts, and finally, it included changes related to testing and verification. 
In another \textit{PR\#174}\footnote{\url{https://github.com/tomhrr/cosh/pull/174}}, the agent added a new feature to the HTTP function so users can view the redirect response content instead of automatically going to the next page. The agent also updated the documentation to show how to use this new option. 

These contributions illustrate how peripheral developers are leveraging agents to handle end-to-end maintenance as well as feature enhancement, reflecting a comprehensive approach to agent delegations.
Conversely, \textbf{\textit{core developers use agents mostly for documentation and testing tasks}} \textit{($42.7\%$ in total)}. In particular, 21.8\% of core developers’ agentic PRs target testing tasks.  
For example, in \textit{PR\#12199}\footnote{\url{https://github.com/keepassxreboot/keepassxc/pull/12199}}, A developer asked the agent to add and test a new search feature that finds all two-factor authentication entries. The agent completed the task and also added tests to make sure the search works correctly, ignoring items without two-factor authentication, handling upper and lower case, and managing exclusion searches properly. 
Core developers also frequently relied on agents for documentation updates (20.9\%). In another \textit{PR\#386}\footnote{\url{https://github.com/numbagg/numbagg/pull/386}}, the developer assigned the agent to write a new documentation file that explains key details for developers, such as what the project is about, how to run commands, how to test the code, and what to check before submitting changes.

Lastly, core developers assign fewer agentic PRs to bug fixing ($-3.8\%$) compared to peripheral developers. This pattern is similar to the studies involving human-authored PRs~\cite{agrawal2018we, mockus2002two}, where peripheral developers act as the ``many eyes'' of the project and most frequently contribute small changes such as bug fixes.

We also investigate how the purposes of agentic PRs depend on the developer group. In particular, we identify two categorical variables - \textit{PR types} and \textit{developer groups}. We use \textit{Chi-Squared} \cite{mchugh2013chi} statistical test to measure the independence between these two variables. We find a statistically significant \textit{p-value} $(p-value \approx 0.0 < 0.0001)$ from our analysis. Thus, there is a statistically significant association between developer group and PR type. This result indicates that the distribution of PR types differs significantly between core and peripheral developers.
\begin{tcolorbox}[
        enhanced,
        colback=teal!5!white,
        colframe=black,
        frame hidden,
        left=2pt,
        right=2pt,
        top=1pt,
        bottom=1pt
    ]
\ding{46} \textbf{RQ\textsubscript{1} Summary:} 
Both groups used agents at comparable rates, but a subgroup of peripheral developers relied on them disproportionately, and core developers achieved higher main/master-branch acceptance rates. Peripheral developers delegated tasks evenly across \textit{bug fixing}, \textit{feature addition}, \textit{documentation}, and \textit{testing}, whereas core developers focused mainly on \textit{testing} and \textit{documentation}.
\end{tcolorbox}


\subsection{Review Dynamics of Agentic PRs (RQ\textsubscript{2})} \label{result-RQ2}
\begin{figure}[!ht]
    \centering
    \includegraphics[width=\linewidth]{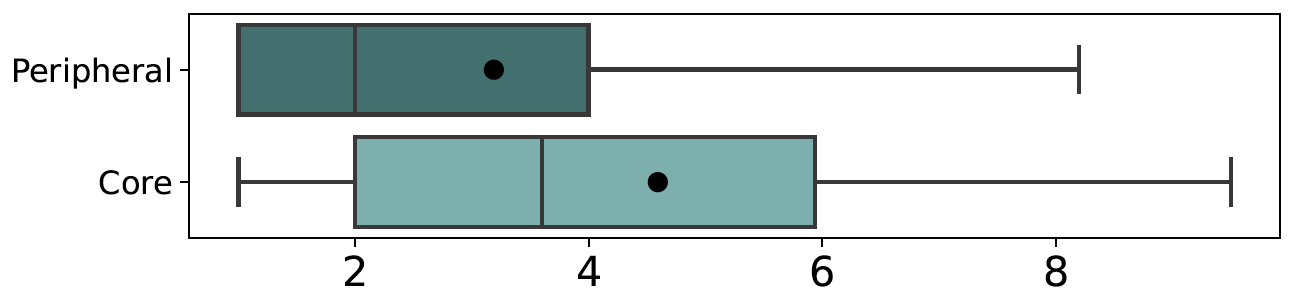}
    \caption{Distribution of reviews comments across peripheral and core developers ($\bullet$ represents mean here)}
    
    \label{fig:review-intensity}
\end{figure}

\begin{figure}[t]
    \centering
    \includegraphics[width=\linewidth]{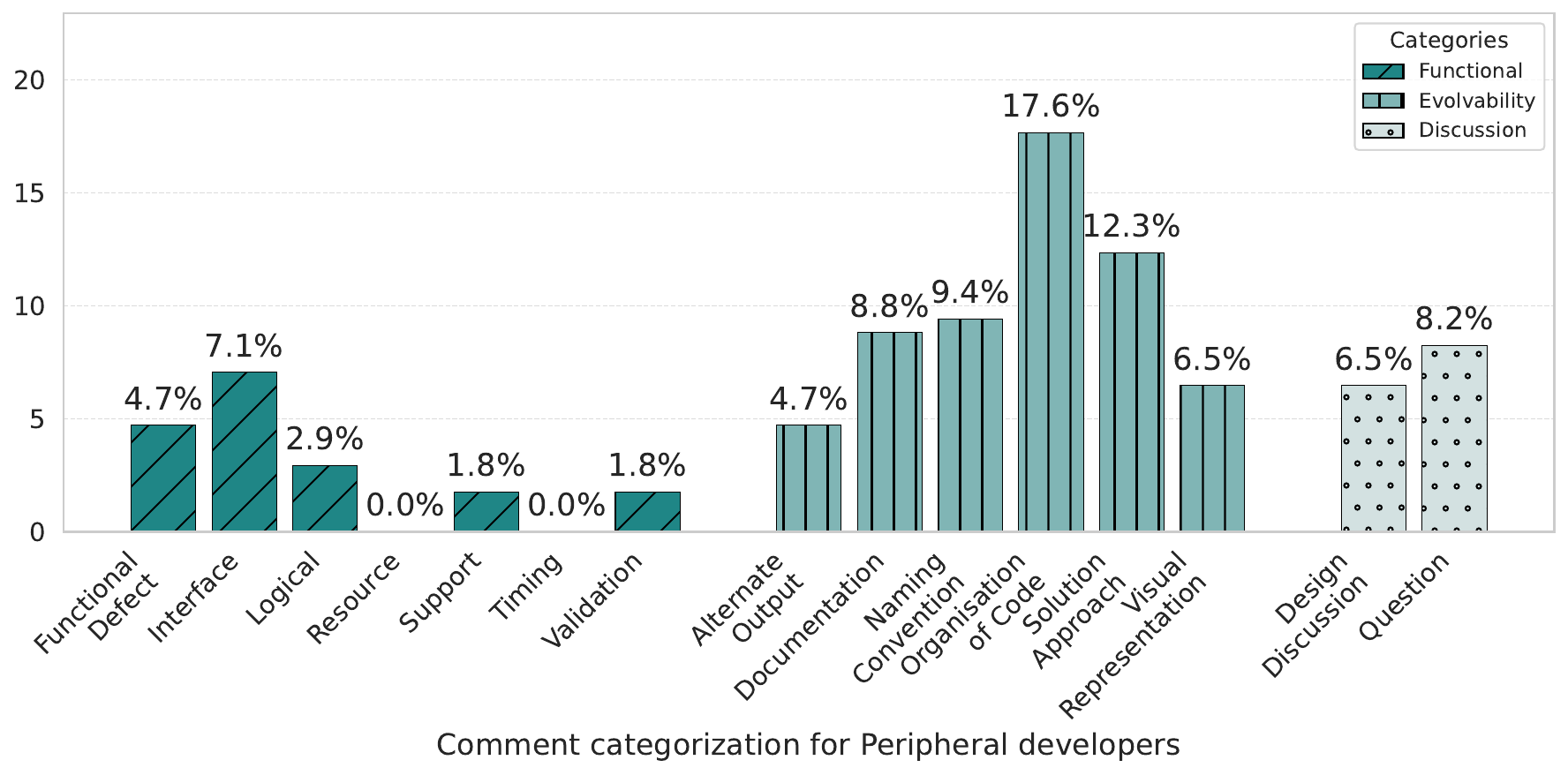}
    \includegraphics[width=\linewidth]{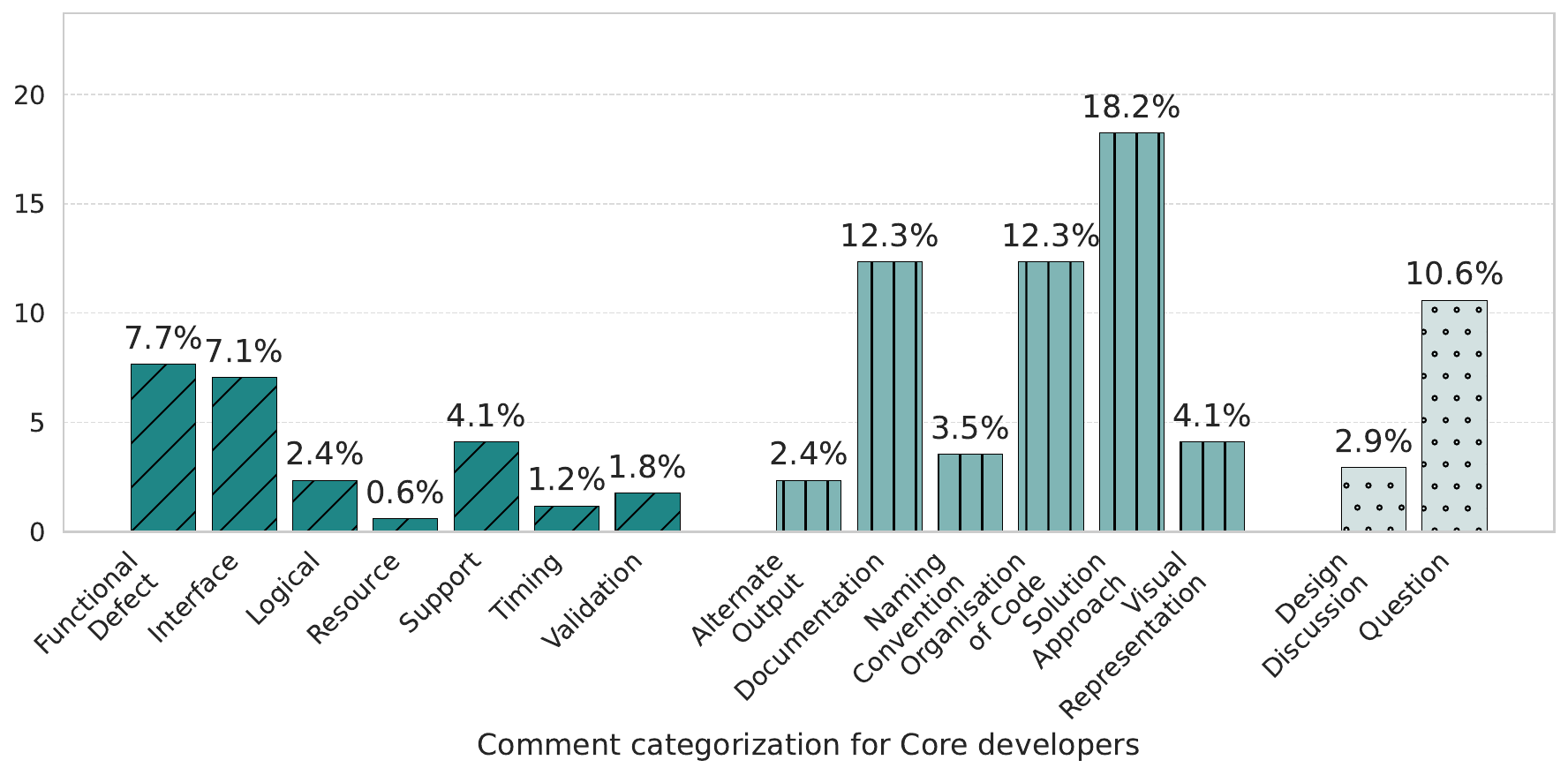}
    \caption{Prevalence of code review comments across developer experience groups}
    \label{fig:review-taxonomy}
    \vspace{-0.5cm}
\end{figure}

Figure ~\ref{fig:review-intensity} summarizes the distribution of review comments across peripheral and core developers. The median number of comments per agentic PR is higher for core developers ($3.6$) compared to peripheral developers ($2.0$). To assess the statistical significance of this difference, we applied the \textit{Mann–Whitney–Wilcoxon} test and found a statistically significant result ($p < 0.001$). The corresponding effect size, measured using \textit{Cliff’s Delta}, was small but non-negligible ($|d| = 0.28$). These findings suggest that core developers engage more thoroughly in review discussions on agentic PRs than peripheral developers, consistent with prior studies on human-authored PRs that identify core contributors as central actors in review~\cite{kerzazi2016can}.
\vspace{-0.5cm}
\vspace{-0.3cm}
Following this, Figure~\ref{fig:review-taxonomy} presents the patterns of issues raised by core and peripheral developers. 

We found that the majority of comments from both groups focused on \textit{evolvability}, aligning with prior observations in human-authored PRs \cite{beller2014modern}. 
Specifically, core developers raised \textit{functional} (24.9\%), \textit{evolvability} (52.8\%), and \textit{discussion}-related issues (14.7\%). Peripheral developers exhibited a similar distribution, with slightly fewer \textit{functional} issues (18.3\%), slightly more \textit{evolvability} issues (59.3\%), and a comparable proportion of \textit{discussion}-related concerns (13.5\%).  Surprisingly, they did not raise any issues related to
\textit{resource} or \textit{timing} types of functional issues in agentic PRs. 

In addition, 
\textbf{\textit{code organization emerged as a prominent concern among peripheral developers}}~\textit{(17.6\%)}.
These review comments mainly focused on suggestions for structural refactoring (e.g., collapse hierarchy), an extra superclass, and an inline function. For example, in \textit{PR\#41332}\footnote{\url{https://github.com/Azure/azure-sdk-for-python/pull/41352}}, the reviewer suggested reorganizing the code by merging one section with another: {\textit{``combine this section with the Message Handling Exceptions section"}}. 
Whereas, 
\textbf{\textit{alternative solutions were frequently observed among core developers' review comments}}~(18.2\%).
Here, developers suggest different approaches, such as alternative algorithms or API choices. For example, in \textit{PR\#22840}\footnote{\url{https://github.com/dotnet/macios/pull/22840}}, the reviewer proposed an alternative implementation by suggesting the use of a more suitable method across the file to improve consistency and correctness:\textit{``Use \texttt{GetIncludedPlatforms} instead of \texttt{GetAllPlatforms} in all cases in this file."}. 
Lastly, both peripheral and core developers frequently seek clarification on design rationale or implementation choices from the agents.
We also use the Chi-Squared test to measure the independence of these two categorical variables - comment category and developer group. However, we did not find any statistically significant association between these two variables.

\begin{tcolorbox}[
        enhanced,
        colback=teal!5!white,
        colframe=black,
        frame hidden,
        left=2pt,
        right=2pt,
        top=1pt,
        bottom=1pt
    ]
\ding{46} \textbf{RQ\textsubscript{2} Summary}:
Core developers engage slightly more in review discussions than peripheral developers. Both groups raise \textit{evolvability} types of review issues largely, with peripheral developers tending to emphasize \textit{code organization} concerns, and core developers more frequently focused on providing \textit{alternative solutions} to the agents.
\end{tcolorbox}

\subsection{Modifications of Agentic PRs (RQ\textsubscript{3})} \label{result-RQ3}

Table~\ref{tab:pr-modification} provides the summary of PRs that were modified by core and peripheral developers after the agent’s initial submission and before final acceptance. 
In most cases, the agent’s contribution was accepted \textit{without modification} with an average of 74.1\% across both developer groups. However, core developers modified a larger share of agentic PRs, representing 28.3\% of their 1,060 accepted PRs, compared to 23.5\% of the 5,759 PRs accepted by peripheral developers. 

\begin{figure}[t]
    \centering
    \begin{subfigure}{0.45\textwidth}
        \centering
        \includegraphics[width=\linewidth]{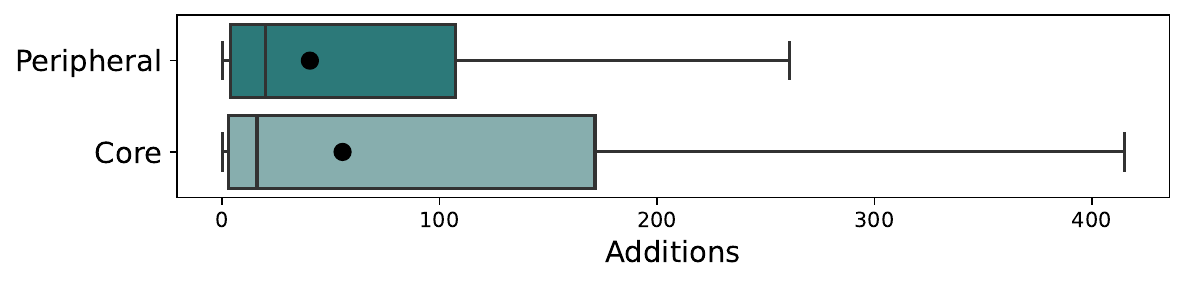}
        \caption{Lines of code added by developers}
        \label{fig:additions}
    \end{subfigure}
    \begin{subfigure}{0.45\textwidth}
        \centering
        \includegraphics[width=\linewidth]{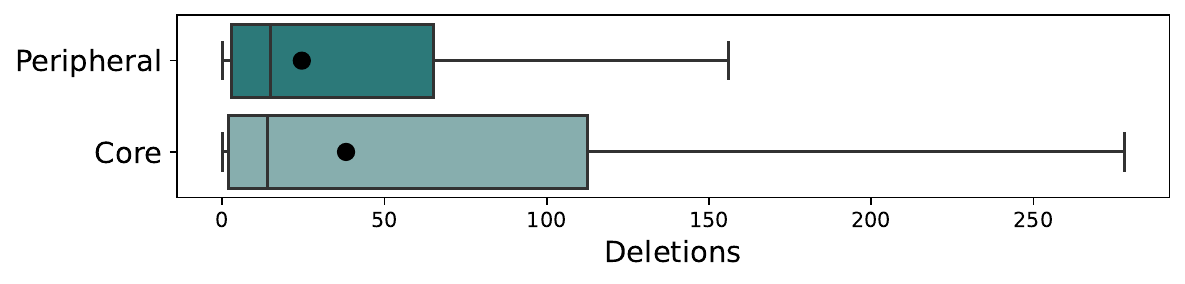}
        \caption{Lines of code deleted by developers}
        \label{fig:deletions}
    \end{subfigure}
    \caption{Added and deleted lines in agentic PR modifications across developer groups ($\bullet$ represents mean here)}
    \label{fig:code-contribution}
    \vspace{-0.5cm}
\end{figure}

In addition, Figure~\ref{fig:code-contribution} shows that the median number of added and deleted lines remains small for both groups, indicating that developers typically make minor adjustments rather than rewriting the agent’s contribution. However, core developers show a higher mean percentage of added lines of code at 55.4\% compared to 40.4\% for peripheral developers. They also have a greater percentage of deleted lines of code, with 38.2\% versus 24.5\% for peripheral developers. The wider spread in core developers' distributions also suggests that, while most modifications are modest, a subset of core developers performs substantially larger edits on agent-generated code. Despite this difference in spread, the Mann–Whitney–Wilcoxon test did not reveal a statistically significant difference between the two groups.

\begin{table}[t]
    \centering
    \caption{Dynamics of Agentic PR modifications}
    \resizebox{1\linewidth}{!}{
    \begin{tabular}{
      >{\raggedright\arraybackslash}p{1.4cm} 
      r
      r
      >{\raggedleft\arraybackslash}p{1.5cm}       
      >{\raggedleft\arraybackslash}p{1cm}     
      >{\raggedleft\arraybackslash}p{1cm}                          
    }
    \toprule
        Developer Category & \#Accepted & \#Modified & Mod. | No Mod. (\%) &  $\overline{\text{LOC}}$ (added) &  $\overline{\text{LOC}}$ (deleted) \\
    \midrule
        Peripheral   &  5759  & 1353 & 23.5 | 76.5  &   40.4   &  24.5  \\
        Core         &   1060  & 300 & 28.3 | 71.5  &   55.4    & 38.2   \\
        
    \bottomrule
    \end{tabular}}
    \label{tab:pr-modification}
\end{table}

\begin{figure}[t]
    \begin{subfigure}{\linewidth}
        \centering
        \includegraphics[width=0.9\linewidth]{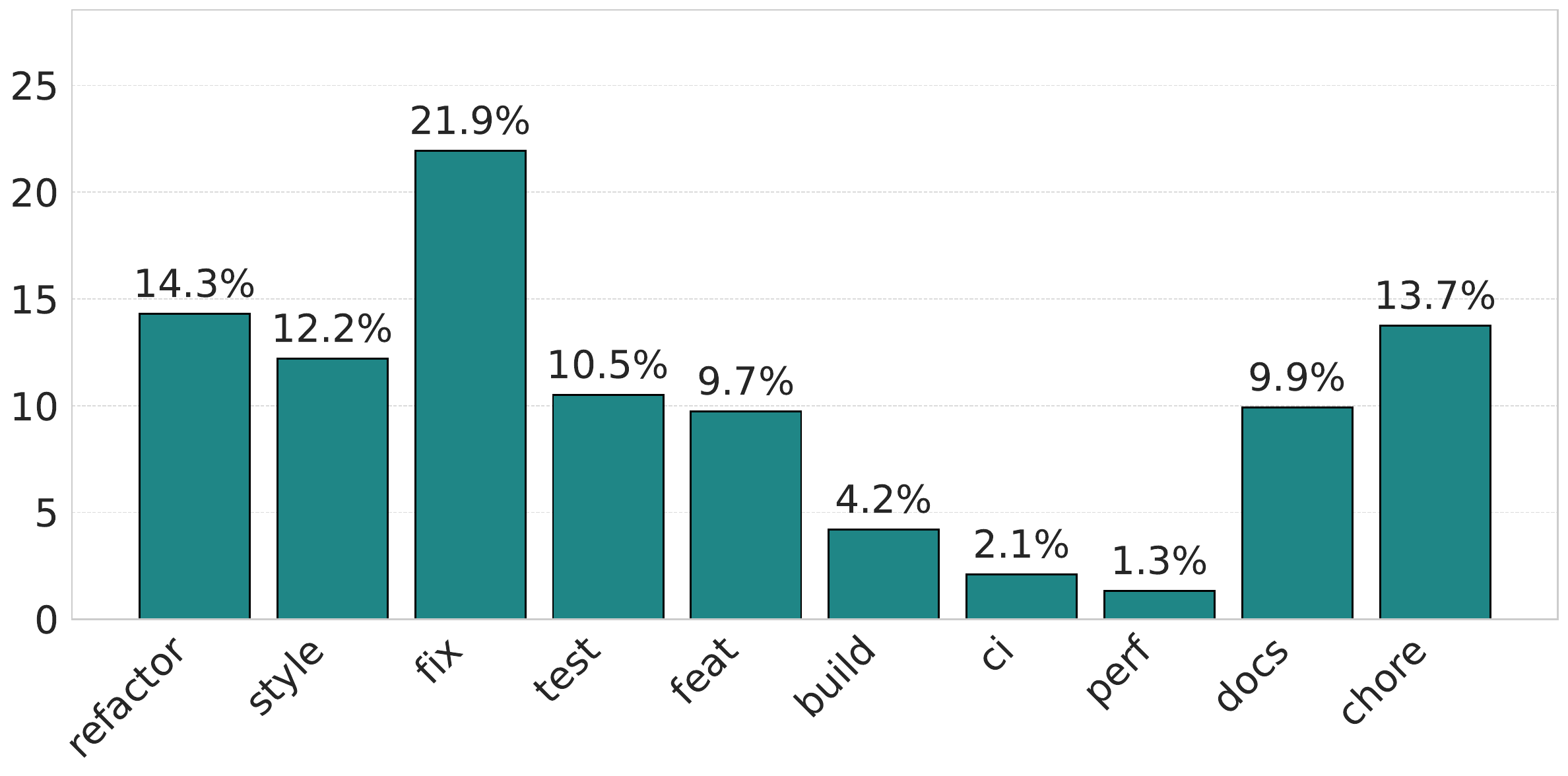}
        \vspace{-0.2cm}
        \caption{Peripheral developers}
        
        \label{fig:peripheral_commit}
    \end{subfigure}
    \begin{subfigure}{\linewidth}
        \centering
        \includegraphics[width=0.9\linewidth]{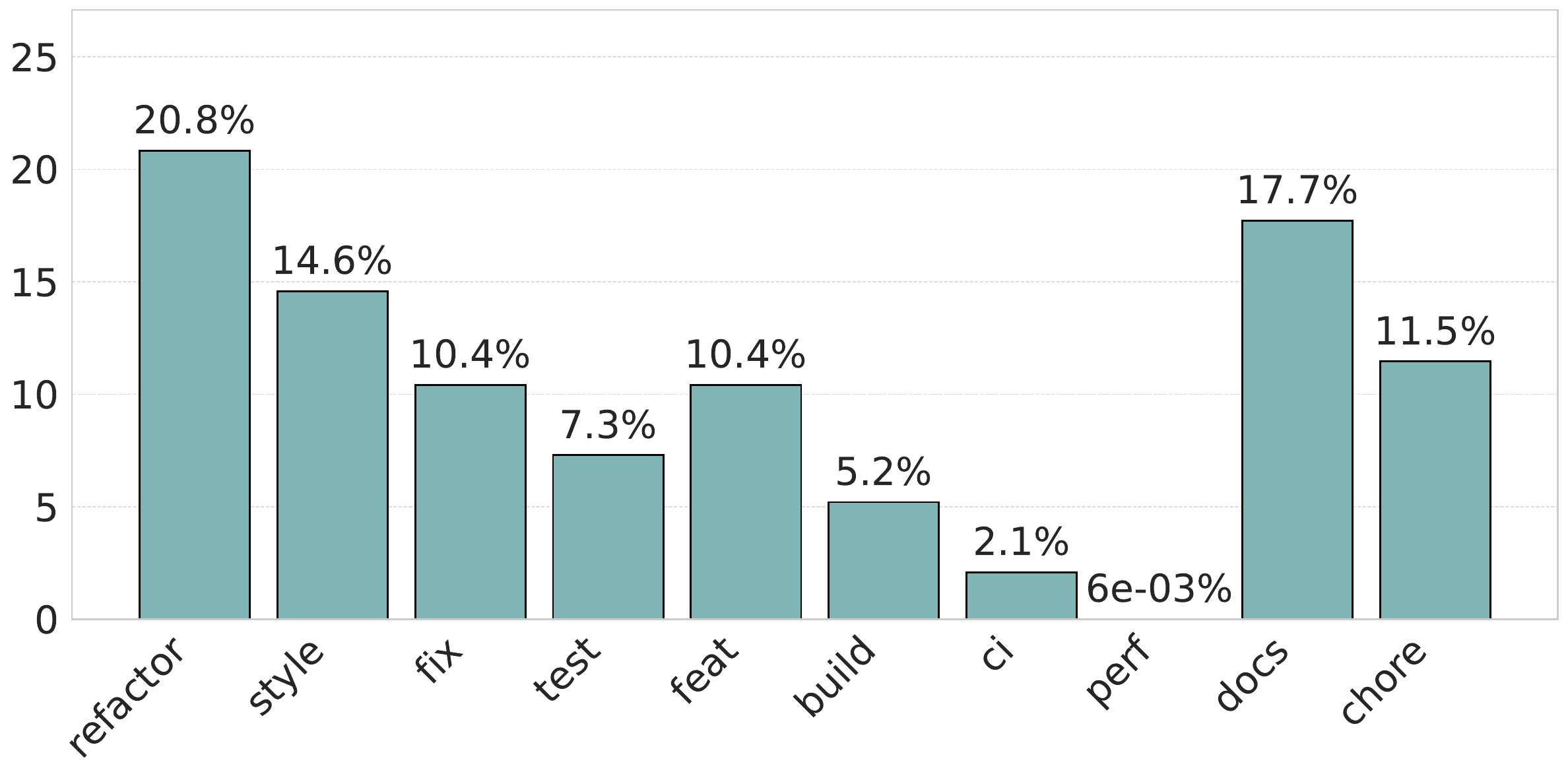}
        \vspace{-0.2cm}
        \caption{Core developers}
        
        \label{fig:core_commit}
    \end{subfigure}
    \caption{Categorization of commits for peripheral and core developers}
    \label{fig:commit_labels}
    \vspace{-0.6cm}
\end{figure}

Subsequently, on our thematic analysis of developers' commit messages, Figure~\ref{fig:commit_labels} summarizes the reasons behind the modifications to agents’ code. 
%
\textbf{\textit{Peripheral contributors frequently modify the agent's code to fix bugs and refactor}}~\textit{(36.2\% in total)}.
For example, in \textit{PR\#4532}\footnote{\url{https://github.com/promptfoo/promptfoo/pull/4532}}, the agent was assigned to check a model’s responses, which produced short, blocked, or unusable answers. To address this, the agent updated the code with an unblocking step that automatically reprocesses such replies. However, when some test cases failed, the developer modified the agent by adding 220 lines and deleting 294 lines of code by fixing type errors, adding safety checks, and fixing the logic based on feedback.
In another example, in \textit{PR\#1532}, the agent reorganized how debugging messages were handled by moving scattered logging code into a shared location for easier reuse and maintenance. After a reviewer noticed a missing version update that could break the code, the developer added 7 lines of code to fix the issue by updating the version number.
 
\textbf{\textit{Core developers mainly refactored the agent’s contribution and improved its documentation}} \textit{(38.5\% in total)}.
%
For instance, in \textit{PR\#2801}\footnote{\url{https://github.com/polkadot-cloud/polkadot-staking-dashboard/pull/2801}}, the agent authored eight commits focusing on big structural tasks, such as moving files, reorganizing components, and cleaning up utilities. These were large-scale, automatic changes that reshaped the project, but the contribution failed to pass all the checks. Then, the core developer made 19 commits (LOC added: 1154 and LOC deleted: 1126) afterward. He fixed type errors, cleaned up code style, adjusted imports and localizatio,n and ensured the CI tests pass. He also reverted one of the Copilot's commits where he manually undid part of the agent's code because it caused issues or was not needed.
In another \textit{PR\#107}\footnote{\url{https://github.com/PythonNest/PyNest/pull/107}}, the agent was asked to add a new feature that helps control who can access different parts of a system. The developer improved the explanation by adding a total of 83 lines, including clear, real-world examples showing how the new feature could be used, making it easier for others to understand and apply.

Lastly, we observe that both developer groups make very \textit{few modifications} to agent-generated code related to build, CI, and performance. This is likely because agents are assigned to these tasks less frequently, as shown in our result Section~\ref{result-RQ1}.
We further applied the Chi-Squared test to measure the independence of these two categorical variables - modification category and developer group. However, the test did not reveal a statistically significant association.
\begin{tcolorbox}[
        enhanced,
        colback=teal!5!white,
        colframe=black,
        frame hidden,
        left=2pt,
        right=2pt,
        top=1pt,
        bottom=1pt
    ]
\ding{46} \textbf{RQ\textsubscript{3} Summary}:
On average, 74.1\% of agentic PRs were accepted without modification. When revisions did occur, core and peripheral developers modified agent-generated code in broadly similar ways. Notably, both groups frequently engaged in \textit{refactoring}, while peripheral developers additionally addressed \textit{bug fixes}, and core developers also focused on \textit{documentation improvements.}
\end{tcolorbox}

\subsection{CI outcomes of Agentic PRs (RQ\textsubscript{4})} \label{result-RQ4}
Table~\ref{tab:checkrun-summary} shows that core contributors are more likely to merge agentic PRs that meet CI requirements compared to peripheral developers. On average, core contributors run slightly more checks per agentic PR than peripheral developers, with averages of 9.3 and 8.1 check runs, respectively.
\begin{table}[th]
\centering
\vspace{-0.3cm}
\caption{Summary of CI check on Agentic PRs}
\label{tab:checkrun-summary}
    \resizebox{0.9\linewidth}{!}{
    \begin{tabular}{
      >{\raggedright\arraybackslash}p{1.4cm} 
      r
      >{\raggedleft\arraybackslash}p{1.2cm}       
      >{\raggedleft\arraybackslash}p{1.2cm}     
      >{\raggedleft\arraybackslash}p{1.4cm}                          
    }
    \toprule
    Developer Category & $\overline{\text{Checks}}$ & Success (\%) & Failure (\%) & No Check (\%) \\
    \midrule
    Peripheral    & 8.1 & 43.1 & 20.0 & 19.1 \\
    Core          & 9.3 & 51.2 & 16.5 & 11.2 \\

    \bottomrule
    \end{tabular}
    }
\end{table}
Additionally, core developers achieve a higher proportion of ``success'' outcomes in all the \texttt{check\_suites} added to the project when compared to peripheral developers, with rates of 51.2\% and 43.1\%, respectively. In contrast, peripheral developers are nearly twice as likely to merge without running any checks on the agentic PR (19.1\% vs. 11.2\%).

For example, in \textit{PR\#34521}\footnote{\url{https://github.com/microsoft/fluentui/pull/34521}}, a peripheral contributor asked Copilot to migrate a set of UI tests. After Copilot’s first commit failed the CI pipeline, the developer replied: \textit{``CI build is failing}.'' Copilot pushed another commit and replied-\textit{``I've resolved the issue […] all tests are now passing.}'', but the CI still failed. The developer then offered minimal guidance, and after two more Copilot attempts, again responded: ``\textit{The tests are failing}.'' After these back and forths, Copilot updated the tests to align with the component’s behavior, and the CI passed. Finally, the developer accepted the PR with two integration checks still failing.  

Whereas, in \textit{PR\#103}\footnote{\url{https://github.com/jamdotdev/jam-dev-utilities/pull/103}}, a core developer tasked Copilot with fixing a minor usability issue (i.e., ensuring that the UI would remember the user's selected view mode instead of resetting after a page reload). Copilot's initial commit successfully addressed the functionality, but when the project's dedicated checks ran, the change failed the \texttt{prettier formatting} check. This indicated that Copilot did not fully adhere to the project's code style guidelines. The core developer then corrected the formatting in a follow-up commit and cleaned up the tests written by Copilot, after which all CI checks passed, and the PR was merged into the \texttt{main} branch.

These results indicate that core developers more consistently adhere to the ``\textit{all checks must pass}'' principle emphasized in CI literature~\cite{soares2022effects, arefeen2019continuous}, whereas peripheral contributors are more willing to merge despite incomplete verification of an agentic PR.

\begin{tcolorbox}[
        enhanced,
        colback=teal!5!white,
        colframe=black,
        frame hidden,
        left=2pt,
        right=2pt,
        top=1pt,
        bottom=1pt
    ]
\ding{46} \textbf{RQ\textsubscript{4} Summary:} 
Peripheral developers are more likely to merge agentic PRs without executing verification checks, whereas core developers more consistently require successful CI results on acceptance.
\end{tcolorbox}

\section{Lessons Learned and Implications}

In this section, we distill six lessons for the community and propose the following implications based on them.

\textbf{What's old is new.} Our findings (Section~\ref{result-RQ1}) show that even though the number of accepted agentic PRs is similar for core and peripheral contributors, core contributors’ agentic PRs are accepted into the main/master branch at a much higher rate. This suggests that socio-technical factors, such as contributor reputation\cite{wang2023fork}, which previously influenced the likelihood of merging human-authored PRs into a project's main branch \cite{ortu2020you, zhou2019fork}, similarly affect agentic PRs. In other words, gatekeeping at the main branch persists even when both groups delegate work to agents; the human social layer still mediates agent outputs. However, we argue that \textit{AI-assisted workflows should be shaped to reduce, not amplify, such long-standing socio-technical inequities} \cite{gao2020deep}. Concretely, the OSS community can establish standardized review rubrics and checklists, ensuring that evaluations are based on the agent's contributions rather than the contributor's socio-technical factors.

\textbf{Agents reducing the time spent on ``toil''.} Our results (Section~\ref{result-RQ1}) show that when developers delegate work to agents, they most often offload toil-heavy tasks (e.g., documentation, testing). Prior studies \cite{kumar2025time, sergeyuk2025using} note that developers often view testing as tedious and would prefer to spend less time on it, and the 2024 Stack Overflow survey~\footnote{https://survey.stackoverflow.co/2024/ai\#developer-tools-ai-next} highlights documentation as a key area where developers expect the greatest growth in AI assistance. These activities are disproportionately seen as work to minimize, positioning them as strong candidates for AI support. Our empirical results confirm that this shift is already underway in real-world OSS projects:\textit{ both core and peripheral contributors predominantly delegate these toil-heavy tasks to agents}, underscoring how AI is being used to absorb low-satisfaction work in practice. 

These findings also point to a practical opportunity for enterprise organizations, especially those with stricter AI usage policies~\cite{schiff2020s}, \textit{to make better use of coding agents for toil}. Companies can support this by educating developers about the strengths and limitations of these tools and by setting clear usage guidelines. As highlighted in Google’s DORA 2025 report\footnote{\url{https://blog.google/technology/developers/dora-report-2025/}}, effective adoption requires more than simply adding AI to the workflow; it also depends on building AI literacy and establishing strong governance practices.

\textbf{Towards cultivating more trust.}  In Section~\ref{result-RQ3}, we show that 74.1\% of agentic PRs are accepted without any developer intervention, consistent with prior findings~\cite{watanabe2025use}. However,  Section~\ref{result-RQ1} reveals that these successes are concentrated in routine tasks such as documentation and testing, whereas developers rarely rely on agents for complex activities like performance tuning or build changes.

Prior research shows that developers benefit when they understand when AI is likely to perform well~\cite{choudhuri2025needs}, and that unclear reliability on complex tasks can miscalibrate trust and reduce productivity~\cite{pearce2025asleep}.
These findings point to a design opportunity for platforms like GitHub:\textit{making agents' competence visible at the moment of use.} For instance, platforms could present confidence indicators based on usage statistics, historical performance, or feedback for similar task types, helping developers decide when to trust, verify, or ignore an agent’s output.

\textbf{Risk of peripherals becoming ``power-users.''} Our results (see Section~\ref{result-RQ1}) indicate that a subset of peripheral contributors drives a significant portion of the agentic flow. They may be using AI to compensate for skill gaps~\cite{crowston2025deskilling} through autonomous agents that submit PRs on their behalf. While this can enable newcomers to contribute more quickly, it raises concerns about an over-reliance on agent-generated PRs, which could hinder learning and create an illusion of progress \cite{brynjolfsson2017artificial}.

Research \cite{masood2022like} shows that newcomers who do not engage deeply with the changes they submit risk struggling to develop long-term skills.  Therefore, to promote sustainable learning, OSS communities could encourage newcomers to start with human-authored contributions before transitioning to more agentic workflows. This approach would ensure that \textit{automation enhances, rather than replaces, developer growth}.

\textbf{Impact of ``Heroes'' on AI-assisted software development.} In Section~\ref{result-RQ3}, we observe that a small subset of core contributors (``heroes'') \cite{agrawal2018we} make substantially larger edits to agent-generated changes before merge. Most of these edits from heroes involve refactoring, which aligns with prior work showing that senior developers focus on maintainability and architecture \cite{Zabardast2023ExploringTRA}. 


Moreover, multiple recent studies also caution that AI-generated code can break quality or security standards, especially in complex projects ~\cite{liu2024refining, fu2025security}. This means core contributors' oversight and careful review are more necessary now before merging agent-generated code to maintain project quality.

\textbf{Core developers remain ``quality gatekeepers":} Our results from Section~\ref{result-RQ4} show that core developers more consistently follow the \textit{all-checks-must-pass} principle and achieve higher “success” rates, while peripheral developers are more likely to merge agentic PRs despite incomplete verification. These patterns align with prior work showing that core developers act as organization quality gatekeepers in agentic contribution as well \cite{gallaba2019improving}. They enforce CI compliance and iterative validation more rigorously than peripheral developers. 

As automation and AI-assistance are increasingly handling testing and build verification, as we also saw in Section~\ref{result-RQ1}, such oversight by core developers becomes pivotal for maintaining accountability and release confidence \cite{Laukkanen2018ComparisonORA}. 
Our findings suggest that teams should promote stricter validation before merging agentic code. For example, adopting \textit{monthly audit} policies and \textit{quality-gate-reviews} led by core developers can safeguard further improve the reliability and maintainability of CI pipelines.

\section{Related Works}
Recent research highlights rapid growth in human–AI collaboration across software engineering tasks such as code search~\cite{liu2024empirical}, code generation~\cite{dakhel2023github}, and automated program repair~\cite{yang2025survey}, with multiple studies demonstrating measurable productivity gains. For instance, Ziegler et al.~\cite{ziegler2024measuring} found that Copilot improves developer productivity across all skill levels, while other studies further confirm that integrating AI-powered tools into development workflows enhances overall efficiency~\cite{banh2025copiloting, takerngsaksiri2025human}. However, subsequent research has revealed limitations in these tools’ effectiveness and their impact on developer cognition~\cite{prather2024widening, vaithilingam2022expectation}. For example, Prather et al.~\cite{prather2024widening} found that generative AI can amplify the metacognitive difficulties of novice programmers, while Vaithilingam et al.~\cite{vaithilingam2022expectation} reported that AI tools do not always improve completion time or success rate and often produce uninterpretable or buggy code.

The impact of AI programming assistants also varies across developer experience levels, with distinct usage patterns~\cite{leinonen2023comparing, aslina2024exploring}. Some studies suggest that less-experienced developers gain greater productivity benefits~\cite{cui2025effects}, while others highlight how AI assistance supports learning by reducing cognitive load and improving usability~\cite{masetty2025enhancing, aslina2024exploring}. In contrast, experienced developers often face challenges such as cognitive interruption and frustration when adapting AI-generated suggestions~\cite{guglielmi2025copilot, prather2023s}. These findings underscore that developer experience fundamentally shapes how AI assistance is used and perceived.

As autonomous coding agents enter the software development landscape, accompanied by bold claims about their potential to replace human developers~\cite{shibu2025anthropic}, research has begun to shift from studying tool capabilities to understanding collaboration dynamics. Terragni et al.~\cite{terragni2025future} envisioned a symbiotic partnership between human developers and AI systems, while Hassan et al.~\cite{ahmed2025can} framed this transition as Software Engineering 3.0, where agents act as intelligent collaborators rather than passive assistants. Yet, empirical evidence remains limited, such as Takerngsaksiri et al.~\cite{takerngsaksiri2025human} found that agents perform well under clearly defined specifications but struggle in agile, conversational workflows. GitHub-based studies~\cite{li2025rise, watanabe2025use} similarly report growing adoption of coding agents, though their PRs are less likely to be accepted for complex tasks. Beyond copilots, developers have also reported challenges with proactive AI systems that act autonomously, often taking unsolicited actions or overwhelming users with excessive information~\cite{erlenhov2020empirical}. Recent work on steering agentic tools~\cite{epperson2025interactive} emphasizes that effective human–agent collaboration requires contextual awareness, transparency, and robust feedback mechanisms.

Together, these findings reveal a paradigm shift, from developers being assisted by AI to collaborating with autonomous coding agents as active teammates. Yet, empirical evidence is lacking on how developers with different experience levels interact with such agents in real-world software development settings. This study addresses that gap by examining how developers of varying experience levels delegate, review, modify, and validate agent-generated code. Understanding these experience-driven collaboration patterns is crucial for designing trustworthy, effective, and sustainable human–AI workflows in modern software development.

\section{Threats to Validity}
Below, we discuss the threats to the validity of our study.

\textbf{External validity.} 
This study focuses on four coding agents, i.e., Claude Code, OpenAI Codex, Copilot, and Cursor. Therefore, the findings may not generalize to all autonomous coding agents or future versions with different interaction capabilities. Moreover, as our analysis is restricted to GitHub-hosted projects, the observed collaboration patterns may not fully represent behaviors in other development environments (e.g., GitLab, Bitbucket, or proprietary enterprise workflows). 

Moreover, we filtered repositories by a minimum threshold of 100 GitHub stars, which may bias our sample toward mature and popular projects. As a result, the findings may not generalize to smaller or emerging repositories, where collaboration practices and agent adoption behaviors could differ.
Lastly, our dataset includes only developers active after agent introduction, potentially limiting generalization to the broader developer population, including multi-repository contributors and newcomers who may interact with agents differently. However, we deliberately applied these filters to ensure that our experience metric reflects preexisting, project-level expertise rather than experience inflated by agent-assisted activity.

\textbf{Internal Validity.}  Our manual coding analyses introduce the subjectiveness of the inspectors to our results. To mitigate this, we follow common practices for all of our manual analyses \cite{krippendorff2018content, lyons2021analysing}. Every disagreement is discussed in a team of two authors until consensus is reached. The coders achieve Cohen’s Kappa scores of 80.3\%–82.5\% for the manual coding tasks.
For PR purpose labeling, we employed GPT-4 for large-scale annotation. While automated classification may introduce labeling errors, we mitigated this by manually validating 364 PRs, achieving a Cohen’s Kappa of 80.3\%, indicating substantial agreement.  

A further threat concerns our measurement of developer experience, which is based on review participation within a single repository. Although this may overlook experience from other projects, we restricted our scope to ensure consistency when examining repository-specific collaboration with coding agents. Our count-based metric may misclassify a small number of contributors as core or peripheral. However, given the large size of both groups, such errors would have a negligible impact, and we do not expect any systematic impact in our results.

Developers may modify agent-generated code through force pushes, which overwrite commit history. Because these edits leave no observable record, some developer interventions may not be captured in our analysis. 
Additionally, we applied the heuristics to distinguish human commits from agent-generated commits according to the current practice at the time of our data collection. Future studies should extend and refine these detection heuristics.
Finally, to calculate CI outcomes, we relied on the \texttt{conclusion} field from the GitHub REST API. This introduces a potential threat to validity because some PRs may be merged even when checks fail, for example, if test failures are unrelated to the submitted changes. As a result, CI status may not always perfectly reflect the true integration of the corresponding agentic PRs.

\textbf{Construct validity.} We use the \textit{Mann-Whitney-Wilcoxon} test, which is a widely used non-parametric test for evaluating the difference between two sample sets. However, the significance level might suffer due to the limited size of the samples. We thus consider the effect size along with $p-value$.
To examine the association between two categorical variables, we also use the \textit{Chi-Squared} test. This statistical test of independence is appropriate when the number of categories is small $(\leq 20)$ \cite{mchugh2013chi}. Thus, threats to construct validity might be mitigated.

\section{Conclusion and Future Work}
This study provides the first empirical evidence of how developers with different experience levels collaborate with autonomous coding agents in real-world software projects. 
Our analysis of 6,819 accepted agentic PRs shows that, although developers exhibit similar acceptance behaviors, their interaction patterns vary by experience. Core developers primarily delegate documentation and testing tasks to agents, engage more deeply in reviews by suggesting alternative solutions, focus on refactoring for maintainability, and achieve higher merge success. In contrast, peripheral developers use agents across broader task categories, emphasize code organization in reviews, mainly fix bugs in agent-generated code, and tend to merge PRs even when verification is incomplete.
These findings highlight that developer experience continues to shape collaboration patterns in the era of agentic software development. Understanding these dynamics can develop better tools, workflows, and training strategies to foster more effective and trustworthy human–AI collaboration.

For future work, we plan to extend this study by examining rejected agentic PRs, exploring longitudinal changes in developer–agent collaboration, and conducting qualitative interviews to better understand how developers perceive trust, accountability, and control when working with autonomous agents. We also aim to investigate how prompt design, feedback mechanisms, and organizational factors influence the success and reliability of agentic contributions.

\section{Acknowledgment}
This research is supported in part by the Natural Sciences and Engineering Research Council of Canada (NSERC) Discovery Grants program, the Canada Foundation for Innovation's John R. Evans Leaders Fund (CFI-JELF), and by the industry-stream NSERC CREATE in Software Analytics Research (SOAR).

\bibliographystyle{ACM-Reference-Format}
\bibliography{MANUSCRIPT}

\end{document}